\pdfoutput=1
\documentclass[letterpaper,twocolumn,10pt]{article}
\usepackage{usenix,epsfig,endnotes}
\usepackage{tabu,longtable,supertabular}
\usepackage[title]{appendix}
\usepackage[english]{babel}
\usepackage[labelfont=bf]{caption} 
\usepackage{epsfig,endnotes,tabu,longtable,supertabular,float,lipsum}
\usepackage{tabularx,url,amsmath,mathtools}
\usepackage[ruled,lined]{algorithm2e}
\usepackage{booktabs,threeparttable}
\usepackage{alltt}
\usepackage{xcolor}
\usepackage{epstopdf}
\usepackage{hyperref}
\usepackage[misc,geometry]{ifsym}
\usepackage{cite}
\usepackage{subfigure}
\usepackage{graphicx}
\usepackage{listings}
\usepackage{flushend}

\makeatletter
\g@addto@macro{\UrlBreaks}{\UrlOrds}
\makeatother

\definecolor{cgray}{gray}{0.90}
\definecolor{cgreen}{rgb}{0,0.6,0}
\definecolor{cmauve}{rgb}{0.58,0,0.82}

\begin{document}

%
%
\title{\Large \bf PerfWeb: How to Violate Web Privacy with Hardware Performance Events}
\hypersetup{pdfauthor={Gulmezoglu et al.},pdftitle={PerfWeb: How to Violate Web Privacy with Hardware Performance Events}}

%
%
\author{
	{\rm Berk Gulmezoglu}\\
	Worcester Polytechnic Institute\\
	bgulmezoglu@wpi.edu
	\and
	{\rm Andreas Zankl}\\
	Fraunhofer AISEC\\
	andreas.zankl@aisec.fraunhofer.de
	\vspace*{1mm}
	\and
	{\rm \hspace*{-3mm}Thomas Eisenbarth\hspace*{3mm}}\\
	\hspace*{-3mm}Worcester Polytechnic Institute\hspace*{3mm}\\
	\hspace*{-3mm}teisenbarth@wpi.edu\hspace*{3mm}
	\and
	{\rm Berk Sunar}\\
	Worcester Polytechnic Institute\\
	sunar@wpi.edu}	

\maketitle

%
%
\begin{abstract}

The browser history reveals highly sensitive information about users, such as financial status, health conditions, or political views. 
Private browsing modes and anonymity networks are consequently important tools to preserve the privacy not only of regular users but in particular of whistleblowers and dissidents. Yet, in this work we show how a malicious application can infer opened websites from Google Chrome in Incognito mode and from Tor Browser by exploiting hardware performance events (HPEs). In particular, we analyze the browsers' microarchitectural footprint with the help of advanced Machine Learning techniques: k-th Nearest Neighbors, Decision Trees, Support Vector Machines, and in contrast to previous literature also Convolutional Neural Networks. We profile 40 different websites, 30 of the top Alexa sites and 10 whistleblowing portals, on two machines featuring an Intel and an ARM processor. By monitoring retired instructions, cache accesses, and bus cycles for at most 5 seconds, we manage to classify the selected websites with a success rate of up to 86.3\%. The results show that hardware performance events can clearly undermine the privacy of web users. We therefore propose mitigation strategies that impede our attacks and still allow legitimate use of HPEs.
\end{abstract}
\section{Introduction}\label{sec:intro}

Web browsers are indispensable components in our lives. They provide access to news and entertainment, and more importantly serve as a platform through which we perform privacy and security sensitive interactions such as online banking, web enabled healthcare, and social networking. Knowing the websites a user is visiting therefore reveals personal and highly sensitive information. To preserve the privacy of users, browsers consequently implement \emph{private browsing} or \emph{incognito} modes, which leave no traces of visited websites. More comprehensive protection is achieved by \emph{Onion routing}, e.g. \emph{Tor}, which protects users against Internet surveillance by obscuring packet routing information. By using a Tor enabled browser users may hide the websites they visit from adversaries monitoring their network communication. This has become indispensable for whistleblowers and dissidents who try to protect their identity against powerful corporations and repressive governments. Besides web browsers, other tools have emerged to mask the identity of the user, e.g. Signal/Redphone, Silent Phone, and Telegram. However, even the installation of such tools can be viewed as subversive action by a repressive regime. In contrast, privacy preserving browsers come pre-installed on many platforms.

While browsers have significantly matured in providing privacy assurances, they are still far from perfect. For instance, an adversary can still infer web browsing activity by exploiting microarchitectural leakages at the hardware level. In 2012, Jana and Shmatikov~\cite{jana2012memento} found that memory footprints of processes are unique and that they can be used to detect opened websites. In 2015, Liu et al.~\cite{liu2015last} demonstrated how the entire last-level cache of a processor can be profiled, which Oren et al.~\cite{OrenEtAl2015} leveraged to infer a small set of opened websites. The key to such inference attacks is that most applications exhibit different execution behavior depending on the input they are processing. They consequently stress the processor hardware in different ways. Whichever application is able to observe these load patterns can learn a great deal of what is being processed in other programs. What eventually enables real-world attacks is that many of the applications we use every day run in the background. Users trust these applications, even though they have little control over what is executed by third-parties. 

In this work, we show that it is feasible for such a third-party application to collect data using hardware performance events (HPEs) and infer private user activity across application boundaries. In particular, we demonstrate that it is possible to infer opened websites, even when users browse in Incognito mode or with the Tor Browser. Such malicious behavior is facilitated in modern operating systems, as HPEs can often be monitored from user space. For the experiments in this work, we use the \texttt{perf} subsystem of the Linux kernel. Since HPE based information is incidental and often noisy, advanced methods for data analysis are needed. The recent advances in Machine Learning (ML) provide us with a powerful tool to classify the complex noisy data in an effective manner. We show that while k-th Nearest Neighbors, Support Vector Machines, and Decision Trees are not sufficient to classify the complex and noisy observed data into a high number of different classes, Convolutional Neural Networks, a Deep Learning technique, can efficiently extract meaningful data even in the presence of severe noise. As a result, we demonstrate that a malicious user space process can infer the web activity of users with very high success rates and in a highly automated fashion.

\medskip
\noindent
{\bf Our Contribution.} In summary, we
\begin{itemize}
\item employ advanced Machine Learning techniques, including Convolutional Neural Networks, and compare their efficiency,

\item use \texttt{perf} to access different types of hardware performance events and combine them to get a better classification rate,

\item cover 40 different websites, including 30 of the top Alexa sites and 10 whistleblowing portals,

\item detect different web pages of a domain to show that fine-grained browser profiling is possible,

\item demonstrate that the attacker does not need to precisely synchronize with the browser, as misalignment is compensated by the ML techniques,

\item show that it suffices to monitor Google Chrome and Tor Browser for at most 5 seconds to classify websites with high accuracy, and

\item outline possible mitigation strategies that impede website inference while still allowing access to performance profiling.
\end{itemize}

The rest of the paper is organized as follows. Section~\ref{sec:background} provides background information and related work for hardware performance events, machine learning techniques, and website fingerprinting. Section~\ref{sec:perf} explains how we measure HPEs, and Section~\ref{sec:scenarios} describes the profiling scenarios. Section~\ref{sec:mlearning} discusses our installments of the ML techniques, before Section~\ref{sec:results} presents the results of our experiments. A further discussion of the results is given in Section~\ref{sec:discussion}. Finally, Section~\ref{sec:countermeasures} describes mitigation strategies and Section~\ref{sec:conclusion} concludes this work.

\section{Background and Related Work}\label{sec:background}

This section provides background information and related work regarding Machine Learning techniques, hardware performance events, and website fingerprinting. Subsequently, we briefly compare our work to previous ones.

\subsection{Machine Learning Techniques}
Machine Learning provides powerful tools to automate the process of understanding and extracting relevant information from noisy observations. All of the techniques we use in this work are \emph{supervised}, meaning that known samples (training set) are used to derive a model that is subsequently employed to classify unknown samples (test set). The success rate of an ML technique in an experiment denotes the percentage of unknown samples that are classified correctly. To reliably determine the success rate, classification is performed multiple times with different training and test sets that are  derived through statistical sampling. This is called cross-validation and is especially useful, if the number of overall samples is low. A brief description of the four ML techniques we use in our experiments is given in the following paragraphs.

\paragraph{\textbf{k-th Nearest Neighbor (kNN).}} The main purpose of kNN is to find the training sample that is closest to a test sample. The Euclidean distance is used to determine how far training and test samples are apart. The smallest distance is taken as the first nearest neighbor and the test sample is marked with the corresponding label~\cite{weinberger2009distance}. As an example, Gong et al.~\cite{gong2010fingerprinting} showed that kNN could be applied to infer websites using remote traffic analysis.
	
\paragraph{\textbf{Decision Tree (DT).}} Decision Trees are used to classify samples by creating branches for the given data features. The general method to decide on the boundaries is to find the feature which gives the best split among the classes. The child branches are then created with other features. While choosing the values for each branch, the entropy is computed to optimize the values. Decision Trees are used by Demme et al.~\cite{DemmeEtAl2013} to detect malware in Intel and ARM processors with HPEs.
	
\paragraph{\textbf{Support Vector Machine (SVM).}} In SVM based learning, input data is converted to a multi-dimensional representation by using mapping functions. Hyperplanes are then created to classify the data. The classification strategy is to find the optimal decision boundaries between classes by increasing the distance between them~\cite{libsvm}. Gulmezoglu et al.~\cite{gulmezoglu2017cache} showed that SVM can be applied in a noisy environment to detect applications that are running in virtual machines on Amazon EC2 cloud.

\paragraph{\textbf{Convolutional Neural Network (CNN).}} Convolutional Neural Networks are one of the most popular Deep Learning techniques that has been proven successful in numerous applications. In contrast to the other ML techniques, CNNs automatically determine important features of the input data. This is achieved by creating nodes between higher and lower dimensional mappings of the input data. The meaningful features are then extracted by finding the optimal functions for each node. When the test data is fed into the CNN, the highest probability is taken as the predicted label~\cite{goodfellow2016deep}. In 2016, Maghrebi et al.~\cite{maghrebi2016breaking} showed that Deep Learning techniques could be applied in side-channel attacks to recover secret information from cryptographic implementations.

\vspace{-2mm}
\subsection{Hardware Performance Events}
The microarchitectures of modern processors implement a large spectrum of performance enhancing features that speed up memory accesses and code execution. As a compromise, performance enhancements introduce input dependent runtimes and weak separation between executing applications. For critical software and mutually untrusted users, this raises severe security and privacy concerns that have been addressed in literature for more than two decades. Kocher ~\cite{kocher1996timing} first describes timing attacks on software implementations of cryptosystems and provides an early anticipation of memory hierarchies, branching units, and variable-time instructions being further exploited. Literature subsequently showed that data and instruction caches~\cite{TromerEtAl2010,AciicmezEtAl2010}, branch prediction units~\cite{AciicmezEtAl2006}, and arithmetic logic units~\cite{AciicmezSeifert2007} can indeed be targeted in attacks. All of them are evidence that the microarchitectural state of a processor contains crucial information about the processes that are executed on it. Hardware performance events are an interface to this state that is implemented on most modern processors. A dedicated piece of hardware, the performance monitoring unit (PMU), is responsible to keep track of microarchitectural events that occur while executing code on the processor. These events include, e.g., instruction retirements, branch mispredictions, and cache references. They provide a comprehensive picture of a processor's runtime behavior and are therefore interesting for adversaries and developers alike. In general, HPEs are useful for application profiling~\cite{AmmonsEtAl1997}, debugging~\cite{YilmazPorter2010,GreathouseEtAl2011}, and even load balancing~\cite{RaoXu2011}. However, the high level of details contained in HPEs also introduces security and privacy issues. Clock cycle events have been recognized as a vital timing source for a large class of cache-based attacks~\cite{ZhangEtAl2016,LippEtAl2016}. In particular, Uhsadel et al.~\cite{UhsadelEtAl2008} demonstrate that cache miss and clock cycle events can be used to mount attacks on software implementations of the Advanced Encryption Standard (AES). Bhattacharya and Mukhopadhyay~\cite{BhattacharyaMukhopadhyay2015} show that branch mispredictions during RSA decryptions reveal the secret exponent because of conditional branches in the multiplication routine during modular exponentiation. In contrast, HPEs have improved our understanding of attacks~\cite{TiriEtAl2007,AticiEtAl2013}, facilitated the evaluation of software components~\cite{ZanklEtAl2016}, and helped to analyze malware samples~\cite{WillemsEtAl2012}. They have also been leveraged to reverse-engineer cache internals on modern processors~\cite{MauriceEtAl2015} and to construct random number generators~\cite{SuciuEtAl2011,MartonEtAl2012}. A large class of previous work is dedicated to the real time detection of attacks and malware infections, a selection of which relies on Machine Learning and related techniques. In particular, naive Bayes~\cite{SinghEtAl2017}, probabilistic Markov models~\cite{KazdagliEtAl2016}, k-Nearest Neighbors~\cite{DemmeEtAl2013}, Decision Trees~\cite{DemmeEtAl2013,KazdagliEtAl2016,SinghEtAl2017}, Random Forests~\cite{DemmeEtAl2013}, Support Vector Machines~\cite{BahadorEtAl2014,TangEtAl2014}, and (Artificial) Neural Networks~\cite{ChiappettaEtAl2016,DemmeEtAl2013,SinghEtAl2017} are studied.

\vspace{-1mm}
\subsection{Website Fingerprinting}
The protection of the browser history is important to ensure the privacy of web users. Yet, literature offers a large spectrum of 
history stealing attacks that allow to recover entries of previously visited websites. Most of them can be launched by malicious web servers and rely on caching~\cite{FeltenSchneider2000} and rendering~\cite{LiangEtAl2014} of website elements, visited URL styles~\cite{JacksonEtAl2006}, and user interactions~\cite{WeinbergEtAl2011}. In addition, attacks have also been demonstrated on the client side in the form of malicious browser extensions~\cite{TerLouwEtAl2008}. If no browsing history is stored, e.g. in private browsing modes, it is still possible to detect websites a user is actively visiting. This is investigated in the field of website fingerprinting, to which we contribute with this work. The following paragraphs discuss different attack vectors for website fingerprinting.

\vspace{-2mm}
\paragraph{\textbf{Network based.}} A significant fraction of website fingerprinting literature is dedicated to network traffic analysis. Attacks typically require an adversary to sniff network communication between the web server and the client machine. Most of the previous works tolerate encrypted traffic, e.g., generated by SSL/TLS or SSH connections, and some even work with anonymized traffic, e.g., routed over the Tor network. To fingerprint and classify websites, previous works have employed a variety of mathematical techniques, many of which are related to Machine Learning. In particular, the Jaccard Index~\cite{SunEtAl2002,SpreitzerEtAl2016}, multinomial naive-Bayes~\cite{HerrmannEtAl2009}, cosine similarity~\cite{ShiMatsuura2009}, Levenshtein distances and related metrics~\cite{LuEtAl2010,CaiEtAl2012,WangGoldberg2013}, k-th Nearest Neighbours~\cite{WangEtAl2014}, Decision Trees~\cite{JuarezEtAl2014}, Random Forests~\cite{HayesDanezis2016}, and Support Vector Machines~\cite{WangGoldberg2013} are studied. 

\paragraph{\textbf{Browser/OS based.}} Website fingerprinting that targets the browser or the underlying operating system typically requires to execute malicious code, e.g. JavaScript, on the client machine. Through this attack vector, Gruss et al.~\cite{GrussEtAl2015} infer opened websites by targeting the memory deduplication feature of modern operating systems and hypervisors. Kim et al.~\cite{KimEtAl2016} exploit the Quota Management API of modern browsers. The authors recover opened websites via storage profiles, which they obtain by continuously reading the remaining space in the temporary storage. Vila and K{\"o}pf~\cite{VilaKoepf2017} employ a similar strategy by timing tasks in shared event loops that handle user interactions on all opened websites. A different approach is proposed by Jana and Shmatikov~\cite{JanaShmatikov2012}, who measure the memory footprint of browsers that is available through the \texttt{procfs} filesystem in Linux. The authors show that different websites exhibit different footprints and subsequently recover opened websites by comparing their footprints to previously recorded ones.

\paragraph{\textbf{Hardware based.}} The third attack vector for website fingerprinting leverages properties of the hardware that runs the web browser. Attacks are mounted by malicious code within the browser, by other processes on the same system, or by an external adversary with physical access to the device. Oren et al.~\cite{OrenEtAl2015} demonstrate that websites exhibit different profiles in the processor cache that can be observed from JavaScript. Hornby~\cite{Hornby2016} also fingerprints websites via the cache, but from another process that is running on the same processor as the web browser. Lee et al.~\cite{LeeEtAl2014} demonstrate that websites can be inferred from rendering traces that are retained in GPUs. The authors obtained these traces with a separate process that is running on the same system as the web browser. Booth~\cite{Booth2015} demonstrates that website fingerprints can also be constructed from the CPU load. The author stresses processor cores via JavaScript and indirectly measures the load of the system that is caused by other opened websites. Experiments are done using kNN classification and dynamic time warping comparisons. Clark et al.~\cite{ClarkEtAl2013} measure the power consumption of laptop and desktop systems and attribute different power profiles to different websites. The authors use Support Vector Machines to classify websites. Yang et al.~\cite{YangEtAl2017} extends this idea to mobile devices that are charged via USB. The authors use Random Forests for website classification.

\subsection{Our Work}
Similar to Hornby~\cite{Hornby2016} and Lee et al.~\cite{LeeEtAl2014}, we assume that a malicious application is running on the same processor as the web browser. In contrast to previous hardware based website fingerprinting, we leverage more than just the processor cache~\cite{OrenEtAl2015} or the processor load~\cite{Booth2015}. To the best of our knowledge, this work is the first that investigates hardware performance events in the context of website fingerprinting. In compliance with the state of the art in this field, we employ supervised Machine Learning techniques in the form of k-Nearest Neighbors, Decision Trees, and Support Vector Machines. While these are recognized instruments for network based fingerprinting, their application to hardware based website inference attacks is still fragmented~\cite{Booth2015,ClarkEtAl2013,YangEtAl2017}. In this work, we directly compare their effectiveness in multiple practical scenarios. In addition, we demonstrate that Deep Learning (in the form of Convolutional Neural Networks) outperforms traditional Machine Learning techniques that are established in both hardware performance event and website fingerprinting literature. To the best of our knowledge, CNNs have not been investigated in neither of these fields before.

\section{Monitoring Hardware Performance Events}\label{sec:perf}

The performance monitoring unit (PMU), which is responsible for counting hardware performance events, implements a set of counters that can each be configured to count events of a certain type. The number of available events is often considerably larger than the number of available counters. Consequently, only a limited number of events can be counted in parallel. In order to measure more events, software layers that use the PMU typically implement time multiplexing. All experiments in this work succeed by measuring only as many events as hardware counters are available, i.e., time multiplexing is not needed. Access to PMUs is typically restricted to privileged, i.e., kernel or system level code, but interfaces exist through which user space applications can gather event counts. On Unix and Linux based operating systems, \texttt{PAPI}~\cite{LKD2015} or \texttt{perf}~\cite{LPM2016} interfaces are commonly implemented. In this work, we focus on the \texttt{perf} interface that is mainly found on Linux systems. Note that this work demonstrates the general feasibility of website fingerprinting with HPEs. Therefore, similar results are also expected on systems with other HPE interfaces.

\subsection{Profiling with Perf}
The \texttt{perf} event monitoring subsystem was added to the Linux kernel in version 2.6.31 and subsequently made available to the user space via the \texttt{perf\_event\_open} system call. Listing~\ref{lst:perfsyscall} shows the system call signature.

\begin{center}
  \begin{minipage}{.48\textwidth}
    \begin{lstlisting}[numbers=none, language=C, caption=\texttt{perf\_event\_open} system call signature~\cite{LPM2016}., label=lst:perfsyscall]
int perf_event_open(struct perf_event_attr *attr,
                      pid_t pid, int cpu,
                      int group_fd,
                      unsigned long flags);
    \end{lstlisting}
  \end{minipage}
\end{center}

The \texttt{perf\_event\_attr} is the main configuration object. It determines the type of event that should be counted and defines a wide range of acquisition properties. We focus only on a very limited number of settings and use zero values for all others. This renders our measurements to be reproducible on a larger number of systems. The \texttt{type} field in \texttt{perf\_event\_attr} specifies the generic event type. As we focus on hardware based events, we only use \texttt{PERF\_TYPE\_HARDWARE} or \texttt{PERF\_TYPE\_HW\_CACHE}. The \texttt{config} field determines the actual event type. The event selection used in this work is given in Section~\ref{sec:scenarios}. In addition, we set the \texttt{exclude\_kernel} option, which avoids counting kernel activity. This improves the applicability of our measurement code, because kernel profiling is prohibited on some systems. Finally, the \texttt{size} field is set to the size of the event attribute struct. The \texttt{pid} and \texttt{cpu} parameters are used to set the scope of the event profiling. In this work, we focus on two profiling scenarios: \emph{process-specific} and \emph{core-wide}. To limit event counting to a single process, \texttt{pid} is set to the process identifier and \texttt{cpu} is set to \texttt{-1}. Subsequently, events are counted only for the given process, but on any processor core. To enable core-wide counting, \texttt{cpu} is set to the core number that should be observed and \texttt{pid} is set to \texttt{-1}. Events are then counted only on one processor core, but for all processes running on it. The \texttt{group\_fd} parameter is used to signal that a selection of events belongs to a group. The \texttt{perf} system then counts all members of a group as a unit. Since this is not a strict requirement for our approach, we omit \texttt{group\_fd} and set it to \texttt{-1}. The \texttt{flags} parameter is used to configure advanced settings including the behavior when spawning new processes and monitoring Linux control groups (cgroups). As none of the settings are relevant to our measurement scenarios, we set \texttt{flags} to zero.

Once \texttt{perf\_event\_open} succeeds, the returned file descriptor can be used to read and reset event counts, and to enable and disable counting. In our measurements, we read event counts using the standard \texttt{read} system call. We found this to yield a sufficiently high sampling frequency and subsequently high success rates during website fingerprinting. On our test systems, the duration of the \texttt{read} system call ranges between 1.5\,$\mu$s and 3.0\,$\mu$s when reading one counter value.

\paragraph{\textbf{Access Control.}} On Linux, access to \texttt{perf} can be configured for user space applications. The access level is specified as an integer value that is stored in \texttt{/proc/sys/kernel/perf\_event\_paranoid} in the \texttt{procfs} filesystem. A negative value grants user space applications full access to performance profiling. If the \texttt{paranoid} level is set to \texttt{0}, comprehensive profiling of the kernel activity is prohibited. A value of \texttt{1} prevents user space applications from core-wide event counting (\texttt{pid\,=\,-1}, \texttt{cpu\,$\geq$\,0}). A \texttt{paranoid} level of \texttt{2} prohibits process-specific event counts while the application gives control to kernel space, e.g., during a system call. Values above \texttt{2} deny event counting even in user space and essentially deactivate \texttt{perf} for user space applications. Note that the \texttt{paranoid} setting is typically overridden by applications started with the \texttt{CAP\_SYS\_ADMIN} capability, e.g., programs started by the root user.

\section{Browser Profiling Scenarios}\label{sec:scenarios}

We investigate the inference of opened websites via HPEs in three distinct scenarios hosted on two Linux test systems. As we are relying on the standardized \texttt{perf\_event\_open} system call of the Linux kernel, there is no need to change the measurement code when switching between systems. The following paragraphs describe each scenario in more detail.

\paragraph{\textbf{1.) Google Chrome on ARM.}} In this scenario, we profile the Google Chrome browser (v55.0.2883) with default options on an ARM Cortex-A53 processor. While the browser loads websites, a malicious user space application is measuring six hardware performance events. In particular, we acquire \texttt{HW\_INSTRUCTIONS}, \texttt{HW\_\-BRANCH\_\-INSTRUCTIONS}, \texttt{HW\_\-CACHE\_\-REFERENCES}, \texttt{L1\_\-DCACHE\_\-LOADS}, \texttt{L1\_\-ICACHE\_\-LOADS}, and \texttt{HW\_\-BUS\_\-CYCLES} events. This selection of events covers instruction retirements, cache accesses, and external memory interfaces. It gives a comprehensive view of the microarchitectural load the browser is putting on the processor. The selected events are measured core-wide, hence including noise from other processes and background activity of the operating system. Since we want to assess the feasibility of core-wide profiling, the browser process is bound to the measured processor core. The events are then measured for five seconds.

\paragraph{\textbf{2.) Google Chrome (Incognito) on Intel.}} In this scenario, we profile Google Chrome in Incognito mode with default options on an Intel i5-2430M processor. The malicious user space application is measuring three hardware performance events, namely \texttt{HW\_\-BRANCH\_\-INSTRUCTIONS}, \texttt{HW\_\-CACHE\_\-REFERENCES}, and \texttt{LLC\_\-LOADS}. In contrast to the ARM scenario, the malicious application acquires process-specific events. Hence, the browser processes float on all processor cores. Since the Intel platform only features three configurable hardware counters, not all of the events measured on ARM can be considered. Compared to the overall retired instructions, we found the retired branch instructions to yield more usable information. As the browser processes are not bound to one core anymore, we substitute events related to the L1 cache with last-level cache loads. In addition, the bus cycle event is omitted, because it is noisier on the Intel platform. The selected events are then measured for one second specifically for the rendering process of the opened website.

\paragraph{\textbf{3.) Tor Browser on Intel.}} In this scenario, we profile the Tor Browser (v6.5.1, based on Firefox v45.8.0) on the same Intel platform as before. In contrast to Chrome, the Tor Browser renders all tabs in one process, which is profiled by the malicious application. While the same performance events are observed, the measurement duration is prolonged. This is because the Tor network introduces significant delays while opening websites.

\paragraph{\textbf{Synchronization.}} None of the scenarios require strict synchronization between the browser and the process of the adversary. Small misalignment is simply passed on to the Machine Learning step. Therefore, we only investigate simple synchronization techniques that can be achieved in practice. For Google Chrome on Intel, the adversary scans the running processes twice per second and checks whether a new rendering process has been spawned. Once a new process is detected, the adversary starts to measure the corresponding process-specific events. The Tor Browser, in contrast, is started freshly for every opened website. The adversary again checks all running processes twice per second and once the Tor Browser is detected, the process-specific profiling is started. This includes additional noise as the browser startup phase is also captured. In the ARM scenario, the measurements are precisely aligned with the start of loading a website. This is used to investigate whether more precise alignment yields better results. Such a trigger signal could be derived from a sudden change or characteristic pattern in the event counts, as the load of the system changes when a website is opened.

\section{Usage of Machine Learning Techniques}\label{sec:mlearning}

After the hardware performance events have been acquired, the measurements for every event are concatenated to create the input data for the Machine Learning techniques. Both training and test sets are normalized to reduce the computation time. Cross-validation is used to obtain reliable success rates. All algorithms are implemented in Matlab 2017a and run on a standard dual-core Intel processor. The training phase of the Convolutional Neural Network is reduced with the help of an NVIDIA Tesla K20 GPU accelerator. Further implementation details of the Machine Learning techniques are discussed in the following paragraphs.

\paragraph{\textbf{k-th Nearest Neighbor (kNN).}} The \textit{fitcknn} command is used to implement kNN and to train our model. By default, the prior probabilities are the respective relative frequencies of the classes in the data, which are initially set to be equal to each other. The Euclidean metric is used to determine the distance between classes.

\paragraph{\textbf{Decision Tree (DT).}} For the Decision Tree, the \textit{fitctree} command is used to train the model. The default values for maximum split is N-1 where N denotes the number of classes. For the training phase, the minimum leaf size is 1 and the minimum parent size is 10.
	
\paragraph{\textbf{Support Vector Machine (SVM).}} We use \texttt{libsvm}~\cite{libsvm} in our experiments to implement multi-class Support Vector Machines. The model is created and trained based on a linear SVM. In general, the type of the SVM can be set to C-SVC or v-SVC. The parameter C is used to regularize the mapping function, whereas the parameter v denotes the upper and lower bound of the fraction of the training error. In our experiments, we chose C-SVC.

\paragraph{\textbf{Convolutional Neural Network (CNN).}} We choose two autoencoders to classify our measurements into N classes. In each autoencoder, different levels of abstraction are learned from the feature vectors and mapped to a lower dimensional space. While the number of layers in the first autoencoder is 100$\,\cdot\,$N, the second autoencoder has 10$\,\cdot\,$N layers. The maximum number of iterations is set to 400 and L2 weight regularization is set to 0.001 for both autoencoders. The last layer is the softmax layer. The training data is trained in a supervised fashion using labels. After the neural network is established and first classification results are obtained, the accuracy of the multilayer network model is improved using backpropagation and repeated training using labeled data. While CNNs have many advantages, the most important disadvantage is their memory demands. When we run out of GPU memory, we downsample the input data to reduce the length of the feature vectors.

\section{Website Profiling Results}\label{sec:results}

In each of the profiling scenarios described in Section~\ref{sec:scenarios}, we monitor HPEs for 30 of the most visited websites according to Alexa~\cite{AlexaInternet2017} (excluding adult sites). This selection is listed in Appendix~\ref{app:urls} (1-30) and used to illustrate the general effectiveness of the Machine Learning techniques to classify websites based on hardware performance events. To demonstrate that also fine-grained website classification is feasible, 10 different sub-pages of the \texttt{Amazon.com} domain are monitored in Google Chrome on Intel. Finally, a selection of whistleblowing websites is measured when visited with the Tor browser. They are also listed in Appendix~\ref{app:urls} (31-40).

\subsection{Google Chrome on ARM}
\begin{figure}[t!]
	\centering
	\includegraphics[width=0.48\textwidth]{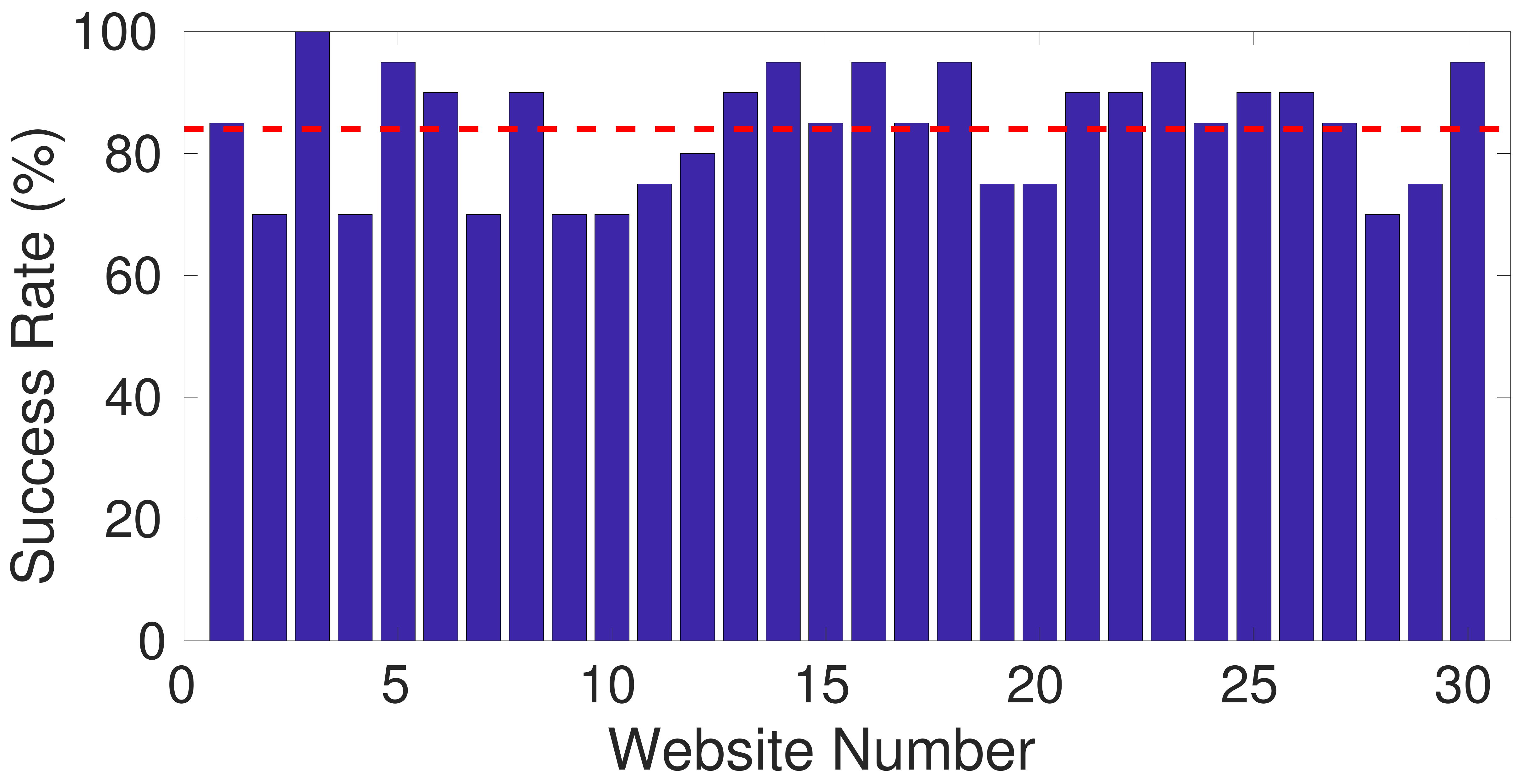}
	\caption{SVM success rates per website for Google Chrome on ARM. The dashed line shows the average classification rate of 84\%.}
	\label{fig:arm_bar}
\end{figure}

For the experiments on ARM, each website is monitored 20 times to train the models. In turn, each measurement consists of 25,000 samples per hardware performance event, which are concatenated for all six events to yield a final measurement size of 150,000 samples. For 30 websites, the total training data size is therefore $90\cdot 10^{6}$ samples. Based on this training set, the success rates after cross-validation are 84\% for linear SVM, 80\% for kNN, and less than 50\% for DT and CNN. The low success rates of DT and CNN indicate that not enough samples have been acquired. Figure~\ref{fig:arm_bar} illustrates the classification rates for each of the visited websites when classified with SVM. Since the number of samples collected in this scenario is small, 10-fold cross-validation is used. The lowest detection rate is 70\%, which shows that core-wide profiling is still feasible even in the presence of noise and background system activity. The average classification rate of 84\% is shown as a dashed line in the figure.

\subsection{Google Chrome (Incognito) on Intel}
For the Google Chrome experiments on Intel, the number of measurements per website is increased to 50. As more samples are acquired, fixed training and test sets are derived instead of using cross-validation. Out of the 50 observations, 40 are used for the training phase whereas 10 are collected to test the derived models. Since each website is monitored for only 1 second, every measurement now consists of 10,000 samples per event. With three observed events, this yields a total training set size of $36\cdot 10^{6}$ and a test set size of $9\cdot 10^{6}$ samples. 

\begin{figure}[t!]
	\centering
	\subfigure[Alexa Top 30]{\includegraphics[width=0.48\textwidth]{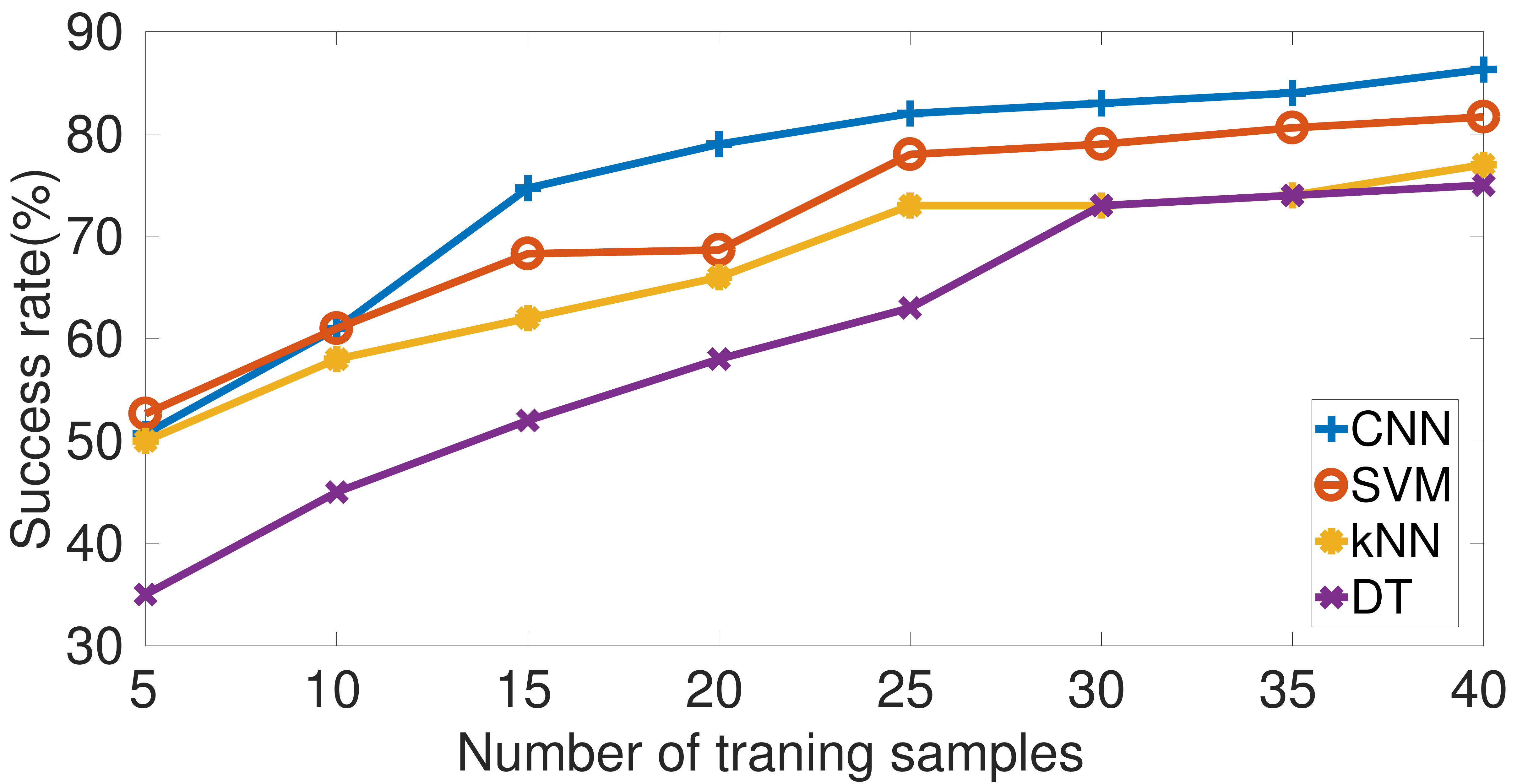}
	\label{fig:chrome_diff}}
	\subfigure[Same Domain Pages]{\includegraphics[width=0.48\textwidth]{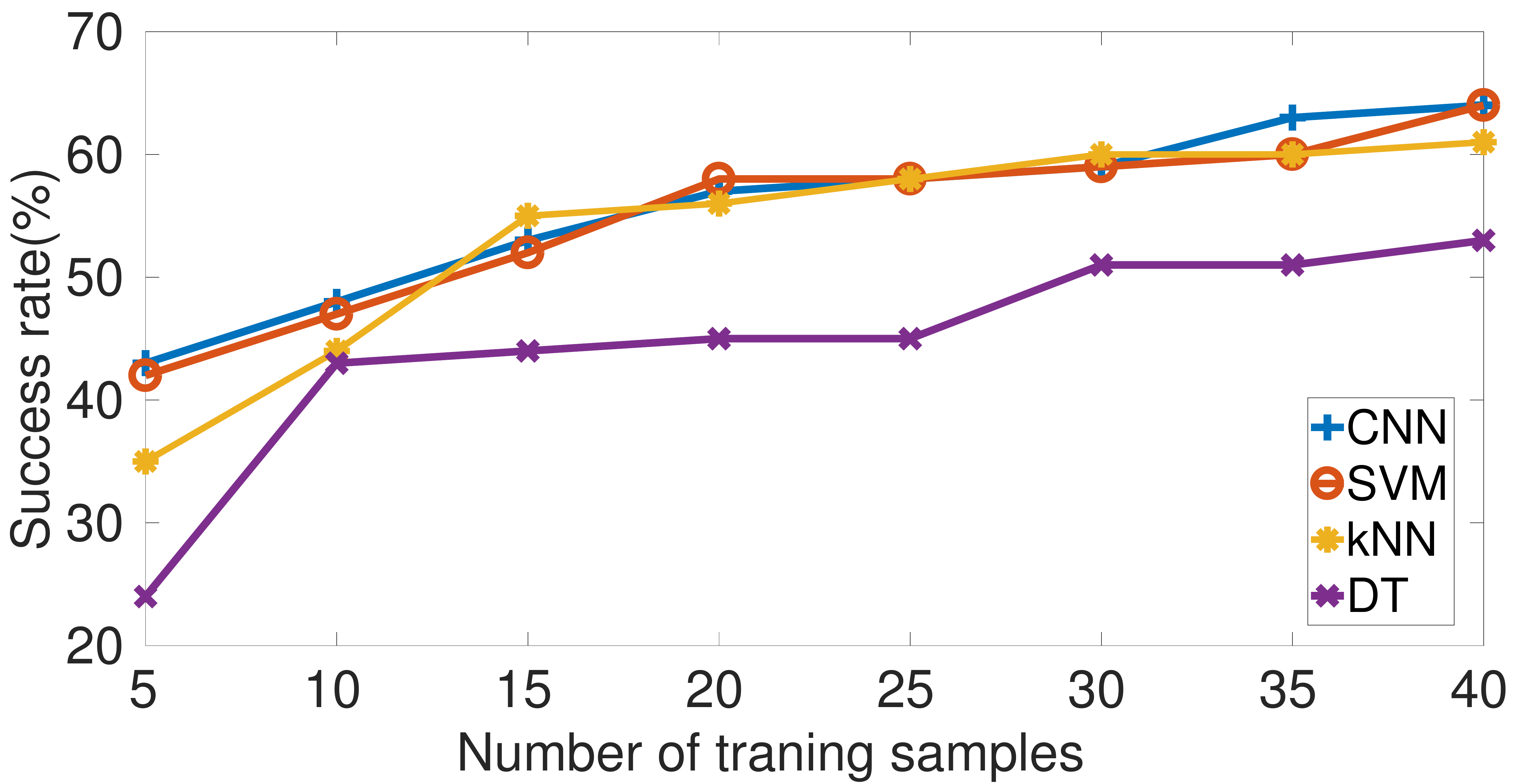}\label{fig:chrome_amazon}}
	\caption{Success rate vs. number of training measurements for Google Chrome (Incognito) and (a) 30 different websites (b) 10 same domain web pages.}
\end{figure}

Figure~\ref{fig:chrome_diff} shows the success rates over an increasing number of training measurements for all Machine Learning techniques and Google Chrome in Incognito mode. Clearly, CNN achieve the highest classification rate, if enough training samples are available. In particular, the success rate for 40 training observations per website is 86.3\%. If the training data size is small, SVM and kNN achieve similar success rates as CNN. Due to the large size of feature vectors in the training and test data, DT gives lower success rates than other ML techniques. Regarding the computation effort, the training phase of CNN takes 2 hours on a GPU and is consequently the longest among the Machine Learning techniques. In contrast, the test phase takes approximately 1 minute for every ML technique.

The second experiment for Google Chrome in Incognito mode on Intel assumes that an adversary has detected a website that the user has visited. Consequently, the attacker tries to infer which page of the website the user is looking at. To illustrate the feasibility of this attack, we selected 10 pages of the \texttt{Amazon.com} domain that display different sections of the online store (kitchen, bedroom, etc.). Naturally, this scenario is more challenging, as the difference between web pages of the same domain is smaller than for entirely different websites. Nevertheless, it is still possible to correctly classify the visited web pages with moderate success. This is illustrated in Figure~\ref{fig:chrome_amazon}. When using CNN and SVM, the success rate is 64\%. kNN yields 60\% success rate, while DT drops to 52\%.

\begin{figure}[t!]
	\centering
	\subfigure[Google Chrome (Incognito)]{\includegraphics[width=0.48\textwidth]{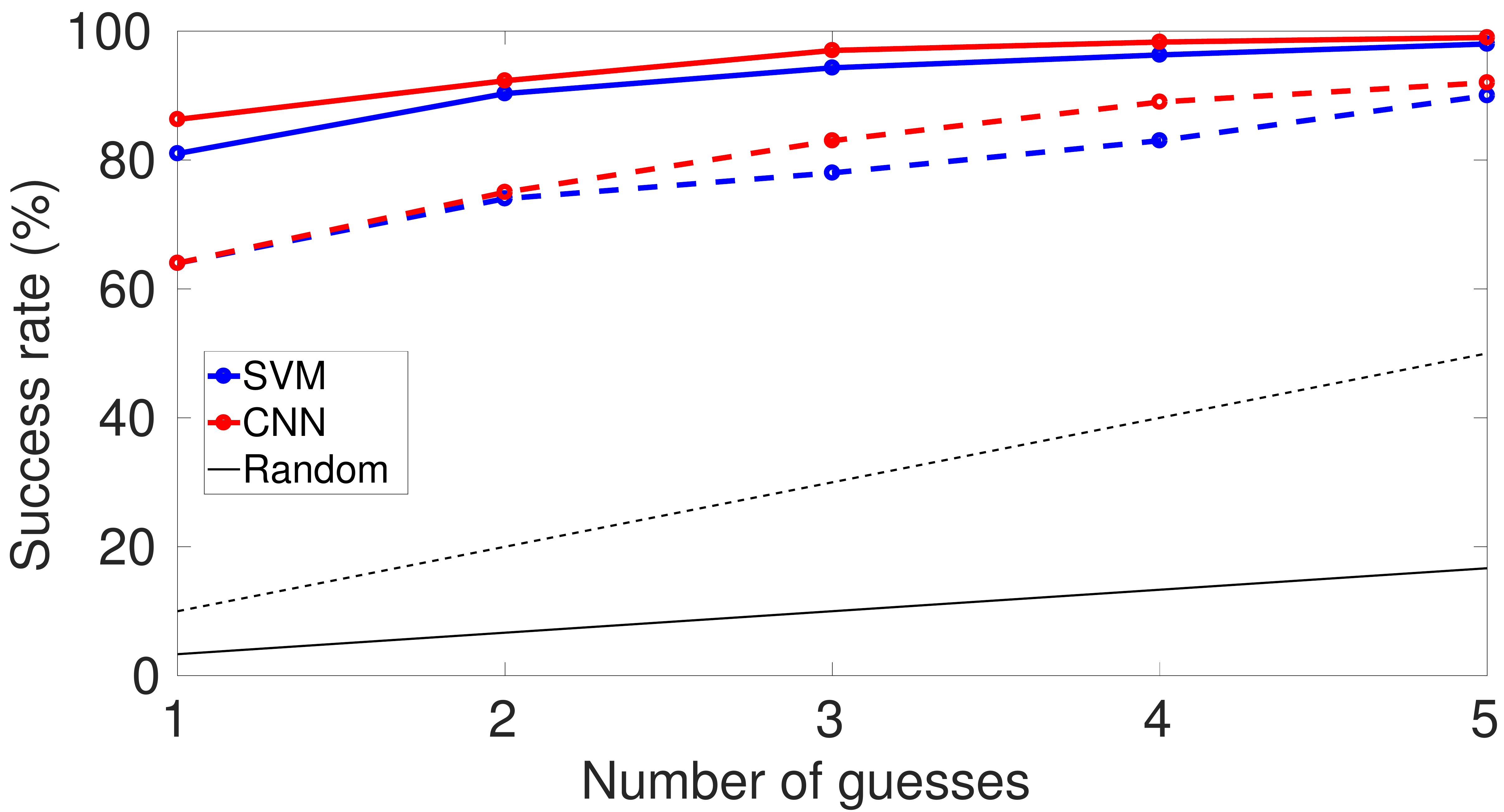}
	\label{fig:guess_incognito}}
	\subfigure[Tor Browser]{\includegraphics[width=0.48\textwidth]{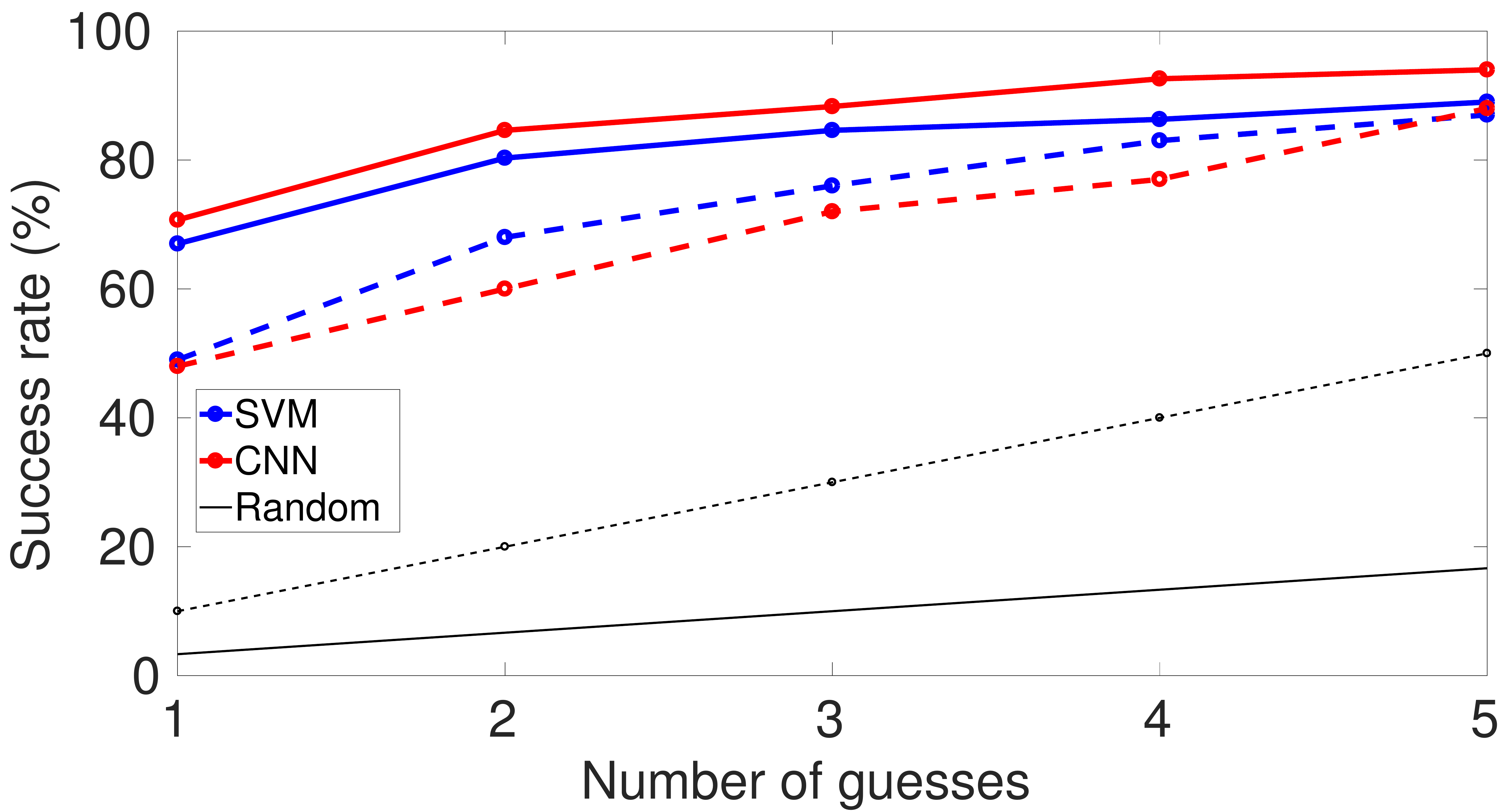}
	\label{fig:guess_tor}}
	\caption{Number of guesses vs. classification rate for (a) Google Chrome (Incognito) and (b) Tor Browser. Solid line represents results for Alexa Top 30, while the dashed line illustrates the same domain results.}
\end{figure}

For CNN and SVM, we also investigate the success rates when the number of guesses are increased. This is shown in Figure~\ref{fig:guess_incognito}. If the first 5 result classes are considered, websites can be detected with 99\% accuracy for SVM and CNN. Similar results are obtained for the same domain experiments, where both CNN and SVM yield 92\% accuracy. Relaxing the number of guesses therefore significantly improves the success rates.

\subsection{Tor Browser on Intel}
For the Tor Browser experiments on Intel, the same events are observed and the same number of measurements are taken for each website. Again, 40 of those measurements are used to construct the training set, while 10 measurement form the test set. As the Tor Browser is monitored for 5 seconds, 50,000 samples are acquired for each event and website. This yields 150,000 samples for one measurement, $180\cdot 10^{6}$ samples for the entire training set, and $45\cdot 10^{6}$ samples for the test set.

\begin{figure}[t!]
	\centering
	\subfigure[Alexa Top 30]{\includegraphics[width=0.48\textwidth]{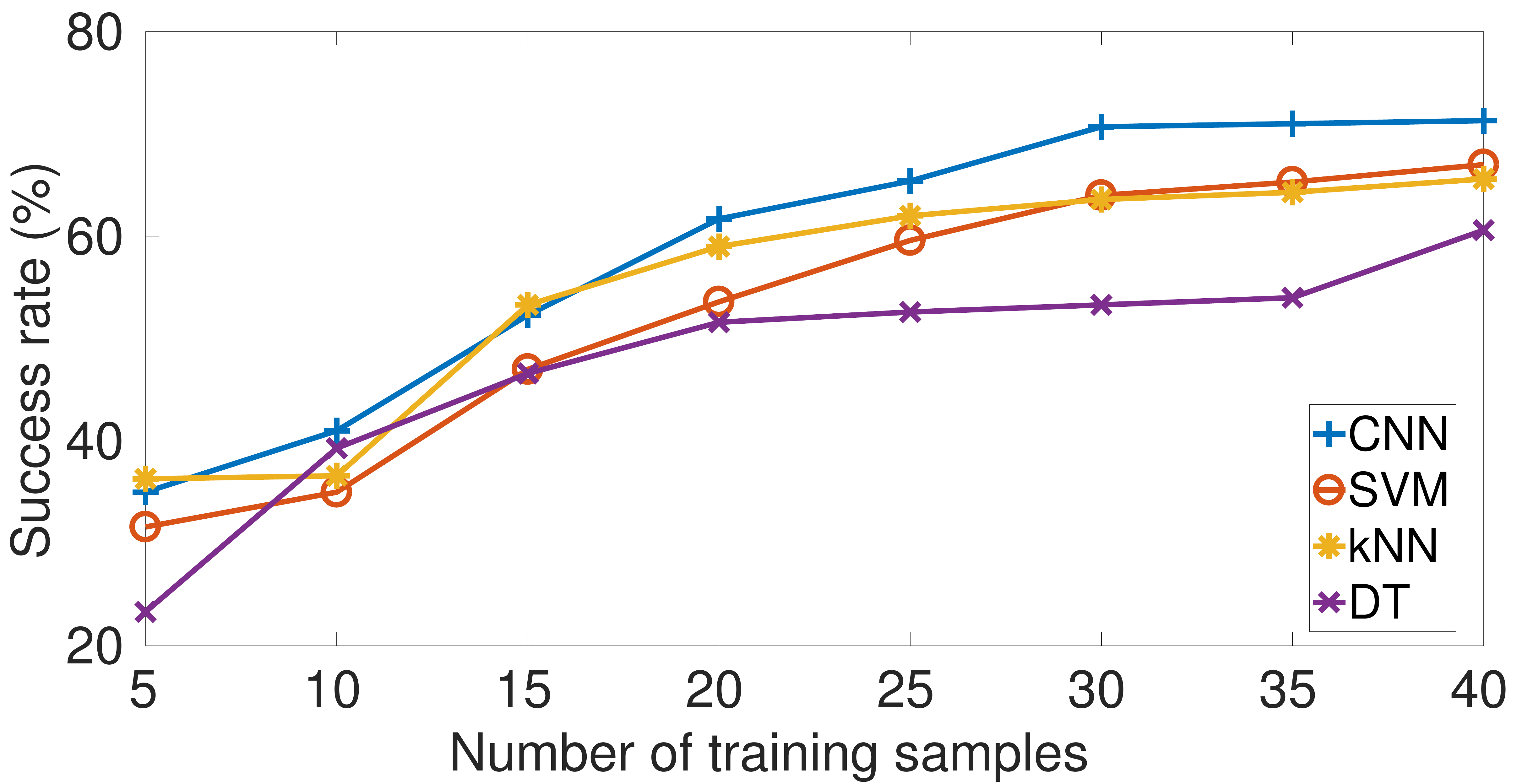}
	\label{fig:tor_diff}}
	\subfigure[Same Domain Pages]{\includegraphics[width=0.48\textwidth]{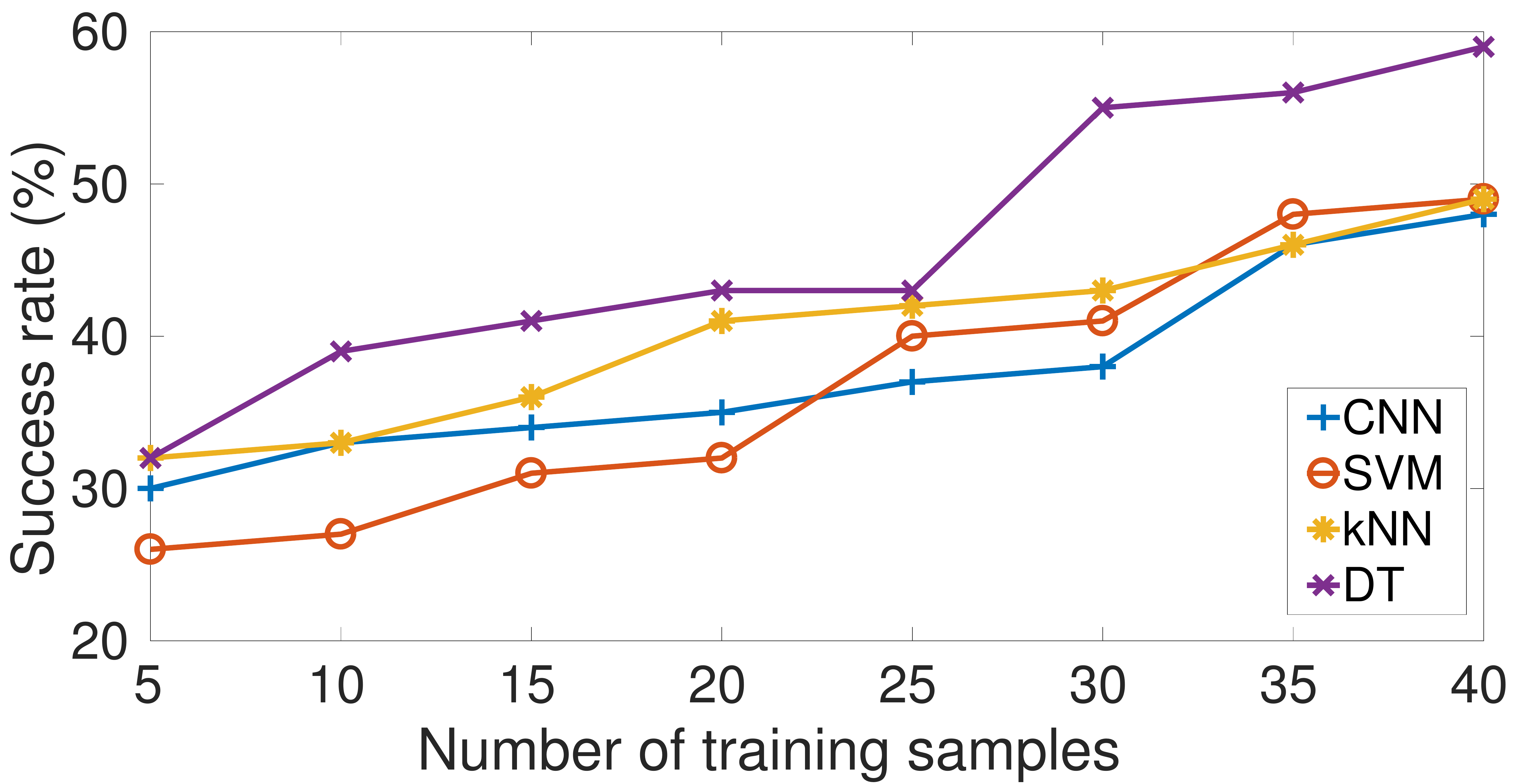}
	\label{fig:tor_amazon}}
	\caption{Success rate vs. number of training measurements for Tor Browser and (a) 30 different websites (b) 10 same domain web pages.}
\end{figure}

Similar to the Google Chrome experiments on Intel, Figure~\ref{fig:tor_diff} shows the success rates over an increasing number of training measurements for all Machine Learning techniques and Tor Browser. CNN yields the highest success rate of 71\%. While SVM and kNN have similar success rates around 66\%, Decision Tree yields a lower accuracy of 60\%. The results show that CNN can handle noisy data and misalignment problems better than other methods, since CNN learns the relations between traces.

The experiment for 10 web pages on \texttt{Amazon.com} is repeated for the Tor Browser and the results are illustrated in Figure~\ref{fig:tor_amazon}. In contrast to the Google Chrome results, Decision Tree yields the highest success rate of 59\%. We believe the reason is the small number of classes that increases the efficiency of DT. The remaining algorithms classify the same domain web pages with a similar success rate of approximately 49\%. Also, Figure~\ref{fig:guess_tor} shows the success rates for CNN and SVM over an increasing number of guesses. While the random selection success rate is around 16\% for 5 guesses, CNN achieves a success rate of 94\%. For the same domain web pages, the success rate of CNN 88\% for 5 guesses. SVM achieves slightly worse results. Slightly increasing the number of guesses thus yields a significant increase in classification success.

\begin{figure}[t!]
	\centering
	\subfigure[All Websites]{\includegraphics[width=0.48\textwidth]{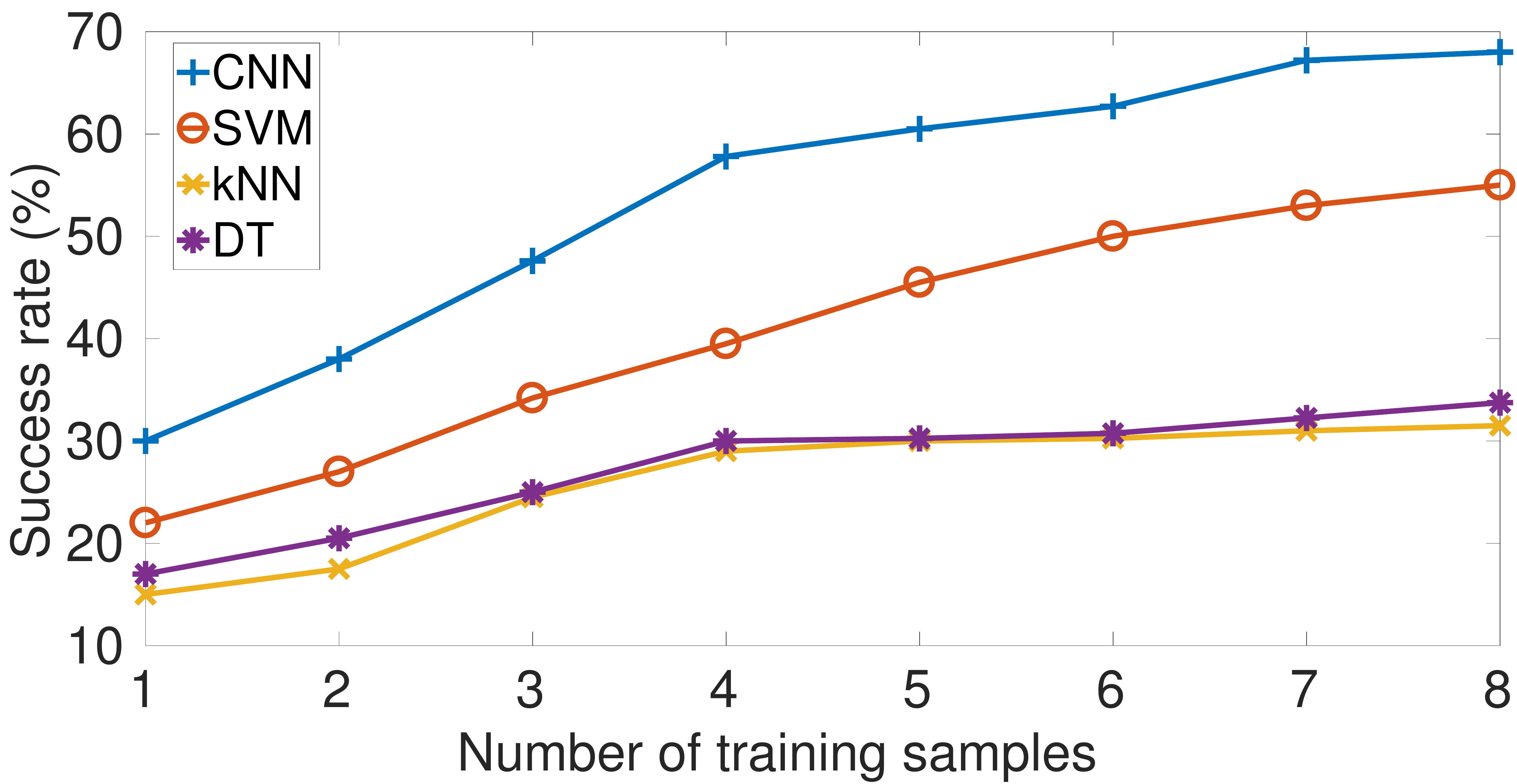}
		\label{fig:tor_all}}
	\subfigure[Whistleblowing and All Websites]{\includegraphics[width=0.48\textwidth]{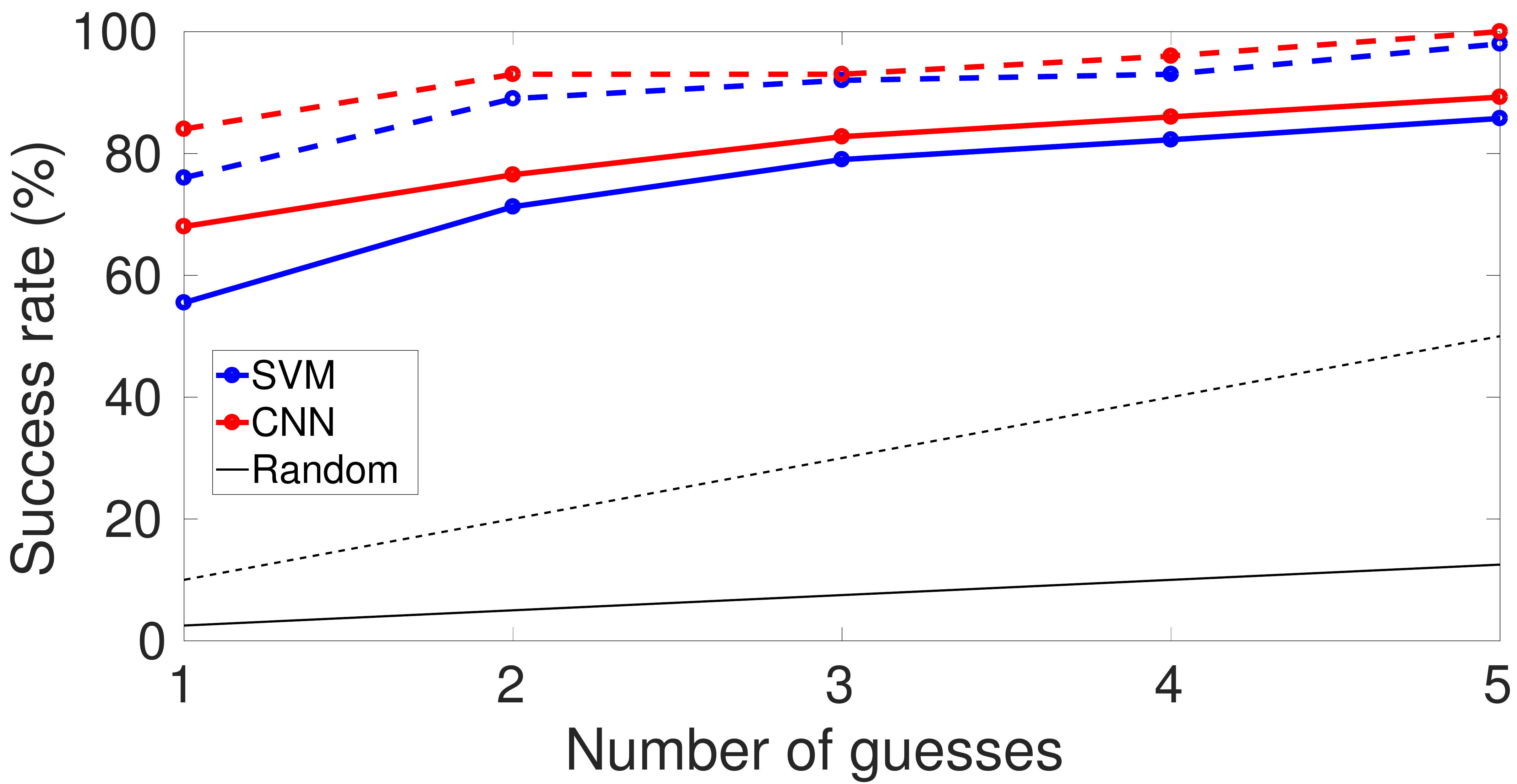}
		\label{fig:whistle_all_guess}}
	\caption{(a) Success rate vs. number of training measurements for Tor Browser and all websites. (b) Number of guesses vs. classification rate for whistleblowing (dashed) and all websites (solid).}
\end{figure}

Finally, we investigate whistleblowing websites, since visiting them anonymously is one of the important reasons to use Tor Browser. For the experiments, we select 10 websites from~\cite{whistle_list}, which are given in Appendix~\ref{app:urls}. In the first step, these whistleblowing websites are classified using all ML techniques. While CNN yields the best classification rate of 84\%, SVM exhibits a success rate of 78\%. In contrast, DT and kNN have lower success rates around 60\%. In the second step, the classification is repeated for all websites considered so far (whistleblowing and Alexa Top 30). Figure~\ref{fig:tor_all} illustrates the success rates for all ML techniques. When classifying 40 websites, CNN yields a success rate of 68\%, while SVM achieves 55\%. In contrast, kNN and DT algorithms cannot classify the websites effectively. When the number of guesses is increased, the success rate improves again. Figure~\ref{fig:whistle_all_guess} shows the classification rates over an increasing number of guesses. If only whistleblowing websites and 5 guesses are considered, CNN yields a success close to 100\%. When all websites are considered, the success rate of CNN is 89.25\%. SVM achieves slightly worse results.

\begin{figure}[t!]
	\centering
	\includegraphics[width=0.48\textwidth]{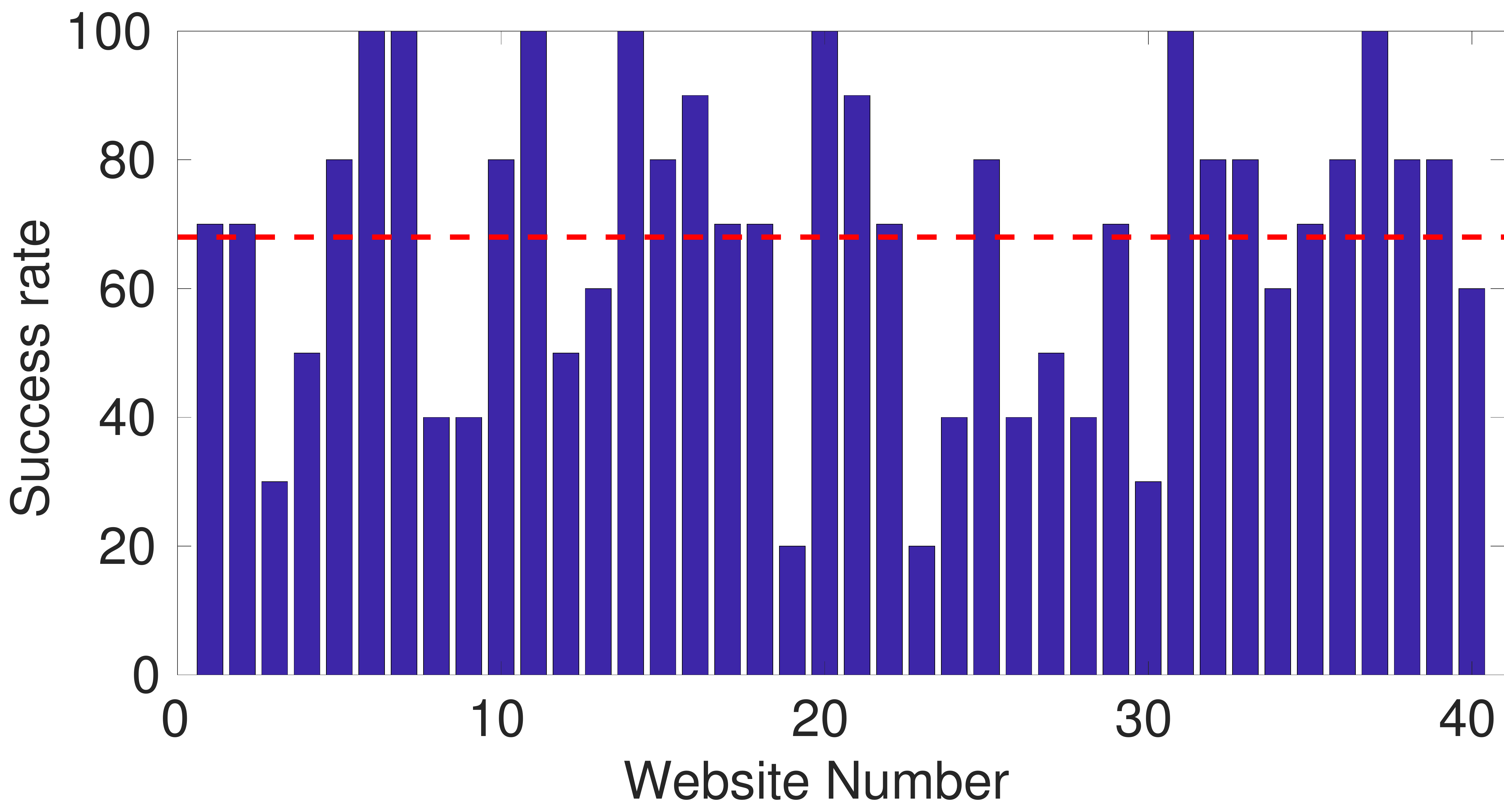}
	\caption{CNN success rates per website for Tor Browser on Intel. The dashed line shows the average classification rate of 68\%.}
	\label{fig:all_bar}
\end{figure}

Individual success rates for CNN are shown in Figure~\ref{fig:all_bar}. The lowest success rate is around 20\% for two websites and seven websites are classified correctly with 100\% accuracy. An interesting observation is that among the 40 websites, the whistleblowing portals are still classified with good success rates. With an average success rate of 68\%, CNN is more capable than other ML techniques to correctly classify websites opened in Tor browser.

\section{Discussion}\label{sec:discussion}

The experiments on ARM were conducted with core-wide measurements, whereas HPEs were acquired in a process-specific fashion on Intel. In general, core-wide acquisition is expected to introduce more noise in the measurements, e.g. from system activity in the background. For process-specific acquisition the activity of the rest of the system does not impede the measurements, as the \texttt{perf} subsystem accumulates event counts only for the specified process. According to the results presented in the previous section, however, both scenarios allow to classify websites with success rates of over 80\% for SVM. Similarly, the precise synchronization on ARM and the approximate process-scanning approach on Intel are both suitable to achieve high classification rates.

Compared to Google Chrome in Incognito mode, the results of the Tor Browser are in general worse. This can be explained with the browser start-up phase, which is always captured for Tor. Also, random network delays introduce jitter in the observations of the website loading. Another adverse effect is the changing geo-location of the Tor exit nodes. Many websites, particularly news sites like New York Times and Yahoo, customize their appearance based on the location of their visitors and therefore introduce additional noise in the measurements.
 
Among the Machine Learning techniques, Convolutional Neural Networks have proven to be the most capable for classifying websites, if enough samples are available. This is the reason why CNNs performed well in Google Chrome and Tor Browser experiments, but not in ARM experiments. CNNs are built for multi-classification of complex structures by extracting meaningful features. On the contrary, SVM and kNN are designed to create hyperplanes to separate space into classes. Since the number of dimensions is high in the experiments, it is difficult to find the best hyperplane for each dimension. Nevertheless, there is a still need for further studies on CNN, since the results could be improved by modifying the parameters, number of layers and neurons. 

In general, the feasibility of website fingerprinting via hardware performance events is not limited to the specific profiling scenarios and test platforms used in our experiments. This is because loading different websites creates different microarchitectural footprints. This a logical consequence of optimized software that is designed to provide best user experience. Therefore, similar results are expected also for other x86 and ARM processors, as well as for other HPE interfaces and web browsers, unless mitigation strategies are implemented.

\section{Countermeasures}\label{sec:countermeasures}

The website inference technique presented in this work has two main requirements. First, websites loaded by a browser exhibit a unique footprint in the microarchitectural state of a processor. Second, this state can be observed via hardware performance events with sufficient precision. Any efforts impacting these two requirements directly affect the reliability, success, or practicality of our approach. The following two paragraphs subsequently discuss such efforts, formulate possible countermeasures, and approximately assess their feasibility.

\paragraph{\textbf{Displaying Websites.}} The first requirement of our classification technique implies that the executed operations during downloading and rendering of website elements are closely related to the type and amount of content displayed on a website. From a more abstract perspective this means that the execution flow and memory accesses of the browser vary for different websites. 
A thorough approach for solving this issue is writing code such that instruction sequences and operand addresses are independent of the input that is processed. While this is reasonable to aspire for security software, it has considerable practical drawbacks in the context of web browsers. First, removing input dependencies almost always impairs performance, because runtime optimizations typically rely on skipping operations and handling special cases differently. As a result, websites take longer to display, which is not in favor of user experience. Second, the larger the code, the more complex it gets to remove input dependencies. For web browsers, at least the code related to networking, storing, and rendering elements must be changed. Given that security critical software has much smaller code bases and still struggles to remove input dependencies in practice~\cite{DoychevKoepf2016}, it is questionable that browser software will successfully implement this in the foreseeable future. If input dependencies cannot be entirely removed, artificial noise can be added to the website loading process. This is, for instance, achieved by introducing random delays between operations or by adding functions that process dummy data instead of real inputs. While this does not solve the underlying problem, it distorts the microarchitectural footprint each website exhibits while being displayed.

\paragraph{\textbf{Observing Events.}} The second requirement is the ability to observe the state of the processor microachitecture with high precision. Since performance monitoring units are dedicated parts of the processor, they cannot simply be removed or permanently deactivated. However, operating systems can block access to them from the software side. On Linux, the kernel can be compiled without the \texttt{perf} subsystem, e.g., by disabling the \texttt{CONFIG\_PERF\_EVENTS} configuration option. Also, the \texttt{perf\_event\_paranoid} file can be set to \texttt{3} or above to disable event counter access from user space. However, blocking or deactivating \texttt{perf} impairs applications that use performance events for legitimate profiling or debugging purposes. If event counting is generally needed, a possible compromise could be more fine-grained profiling restrictions, such that processes can only count events caused by themselves. Profiling any other process is prohibited, even if it belongs to the same user. While this requires changes to the \texttt{perf} interface, it provides legitimate applications access to profiling and at the same time impairs the fingerprinting technique presented in this work. This profiling restriction could be conveniently added as a dedicated setting in the \texttt{perf\_event\_paranoid} configuration file. An alternative solution is to lower the measurement precision of hardware performance events. This can, for instance, be achieved by artificially adding a certain level of noise to the event counts while retaining a sufficiently high signal-to-noise ratio, or by reducing sampling frequencies with which applications can acquire event counts. Yet again, this would also affect benign applications. A possible solution is to detect malicious programs and then only degrade their observations. However, the presented measurement approach behaves identically to legitimate applications and does not rely on exotic operations or measurement settings.

\section{Conclusion}\label{sec:conclusion}

When websites are loaded in the browser, they stress the underlying hardware in a distinct pattern that is closely related to the contents of the website. This pattern is reflected in the microarchitectural state of the processor that executes the browser, which can be observed with high precision by counting hardware performance events. Since these events can be legitimately measured by user space applications, it is feasible to infer opened websites via performance event measurements. We demonstrated this by utilizing Machine Learning techniques, achieving high recognition rates even in the presence of background noise, trace misalignment, and varying network delays. In addition, the results show that CNN is able to obtain better classification rates from high number of classes in the presence of noise. By applying CNN, the whistleblowing websites are classified with 79\% accuracy among 40 websites while the overall classification rate increases up to 89.25\% with 5 guesses in Tor browser.

%
%
\section*{Acknowledgments}
This work is supported by the National Science Foundation, under grants CNS-1618837 and CNS-1314770.

%
%
{\footnotesize \bibliographystyle{acm}
\bibliography{references}}

%
%
\begin{appendix}
\section{List of Profiled Websites}\label{app:urls}

\noindent
\vspace{-5mm}
\begin{table}[h]
	\centering
	\caption{Websites profiled in this work. Entries 1-30 are taken from the top websites listed by Alexa~\cite{AlexaInternet2017}, while URLs 31-40 are a selection of whistleblowing portals.}
	\label{table:websites}
	\begin{tabular}{@{}ll@{}}
		\toprule
		\multicolumn{2}{c}{Website Number and URL}      \\ \midrule
		1) Netflix.com  & 21) Office.com   \\
		2) Amazon.com   & 22) Microsoftonline.com    \\
		3) Facebook.com & 23) Chase.com      \\
		4) Google.com   & 24) Nytimes.com\\
		5) Yahoo.com    & 25) Blogspot.com\\
		6) Youtube.com  & 26) Paypal.com\\
		7) Wikipedia.org& 27) Imdb.com  \\
		8) Reddit.com   & 28) Wordpress.com\\
		9) Twitter.com  & 29) Espn.com\\
		10) Ebay.com    & 30) Wikia.com\\
		11) Linkedin.com& 31) Wikileaks.org \\
		12) Diply.com   & 32) Aljazeera.com/investigations\\
		13) Instagram.com& 33) Balkanleaks.eu \\
		14) Live.com    & 34) Unileaks.org\\
		15) Bing.com   & 35) Globaleaks.com\\
		16) Imgur.com  & 36) Liveleak.com\\
		17) Ntd.tv    & 37) Globalwitness.org\\
		18) Cnn.com    & 38) Wikispooks.com\\
		19) Pinterest.com & 39) Officeleaks.com \\
		20) Tumblr.com & 40) Publeaks.nl\\
	\end{tabular}
\end{table}
\end{appendix}

\end{document}